\documentclass{iaus}
\usepackage{graphicx}

\title
[On the self-consistent physical parameters of LMC intermediate-age clusters]
{On the self-consistent physical parameters of LMC intermediate-age clusters}

\author[Leandro Kerber \& Bas\'\i lio Santiago]
{Leandro Kerber$^{1,2}$
 \and Bas\'\i lio Santiago$^3$}

\affiliation{$^1$IAG/USP, S\~ao Paulo, Brazil \\ 
email: {\tt kerber@astro.iag.usp.br}
\\[\affilskip]
$^2$INAF-OAPd, Padova, Italy
\\[\affilskip]
$^3$IF/UFRGS, Porto Alegre, Brazil}

\pubyear{2008}
\volume{256}  
\pagerange{119--126}
\jname{The Magellanic System: Stars, Gas, and Galaxies}
\editors{Jacco Th. van Loon \& Joana M. Oliveira, eds.}
\begin{document}

\maketitle

\begin{abstract}
The LMC clusters with similar ages to the Milky Way open clusters
are in general more metal-poor and more populous than the 
latter, being located close enough to allow their stellar content to
be well resolved.   
Therefore, they are unique templates of simple stellar 
population (SSP), being crucial to calibrate models describing 
the integral light as well 
as to test the stellar evolution theory.
With this in mind we analyzed HST/WFPC2 (V, B--V) colour-magnitude diagrams 
(CMDs) of 15 populous LMC clusters with ages between 
$\sim$0.3 Gyr and $\sim$4 Gyr
using different stellar evolutionary models. 
Following the approach described by \cite[Kerber, Santiago \& Brocato (2007)]
{KSB07}, we determined 
accurate and self-consistent physical parameters (age, metallicity, 
distance modulus and reddening) for each cluster by comparing the 
observed CMDs with synthetic ones generated using isochrones from 
the PEL and BaSTI libraries.
These determinations were made by means of simultaneous 
statistical comparison of the main-sequence fiducial line and the red 
clump position, offering objective and robust criteria to select 
the best models.
We compared these results with the ones obtained by 
\cite[Kerber, Santiago \& Brocato~(2007)]{KSB07} using the Padova isochrones. 
This revealed that there are significant 
trends in the physical parameters due to the choice of stellar 
evolutionary model and treatment of convective core overshooting.
In general, models  that incorporate overshooting 
presented more reliable results than those that do not.
Furthermore, the Padova models fitted better the data than the PEL and
BaSTI models.
Comparisons with the results found in the literature demonstrated 
that our derived metallicities are in good agreement with 
the ones from the spectroscopy of red giants. 
We also confirmed that, independent of the adopted stellar evolutionary
library, the recovered 3D distribution for these clusters is
consistent with a thick disk roughly aligned with the LMC disk as 
defined by field stars.
Finally, we also provide new estimates of distance modulus to the 
LMC center, that are marginally consistent with the canonical 
value of 18.50. 
\end{abstract}

\firstsection 
\section{Introduction}

The LMC contains a rich system of stellar clusters, with more than
three thousand cataloged objects (\cite[Bica \etal\ 2008]{Bica_etal08}) and 
covering ages from few Myr to about 13 Gyr. 
There are about one hundred that can be considered
as populous ones ($> 10^{5}$ stars), which offer the opportunity 
to recover the age-metallicity relation for the LMC by means of 
accurate age (from CMD analysis) and 
metallicity (from spectroscopy analysis) determinations. 
Furthermore, the objects with ages between $\sim$ 0.3 and 4 Gyr -- 
the intermediate-age LMC clusters (IACs) -- 
are more metal poor than the open clusters in the Milky Way, 
being therefore fundamental pieces in the local universe 
to calibrate integrated light models as well as 
to test the evolutionary models in the sub-solar metallicity regime. 

Taking the advantage of the superior photometric quality of the HST
and the intrinsic large stellar statistics for the populous IACs 
we have applied a statistical method to recover accurate 
physical parameters -- not only age, but also metallicity, 
distance and reddening -- for these objects in a self-consistent way. 
By self-consistency we mean the ability to simultaneously 
infer these parameters from the same data-set without any prior 
assumptions about any of them.
The method, presented by 
\cite[Kerber, Santiago \& Brocato~(2007)~(hereafter KSB07)]{KSB07}
using the Padova isochrones (\cite[Girardi \etal\ 2002]{Girardi_etal02}), 
joins CMD modelling and statistical analysis to objectively determine which 
are the synthetic CMDs that best reproduce the observed ones. 
In the present work we expanded this analysis to two other   
stellar evolutionary libraries: 
PEL (or Pisa, \cite[Castellani \etal\ 2003]{Castellani_etal03}) and 
BaSTI (or Teramo, \cite[Pietrinferni \etal\ 2004]{Pietrinferni_etal04}).
This allowed us to quantify how the recovered physical parameters 
depend on the adoption of different stellar evolutionary libraries, 
including the treatment for the convective core overshooting.
Furthermore, since we are determining the individual distance to each
cluster, we could also probe the three dimensional 
distribution of these 
clusters, which seems to be roughly aligned with LMC disk 
(\cite[KSB07]{KSB07}; \cite[Grocholski \etal\ 2007]{Grocholski_etal07}), 
and to obtain new determinations of distance modulus to the LMC centre.      

\section{CMDs: data vs. model}

\begin{figure}[b]
\begin{center}
 \includegraphics[width=10.0cm]{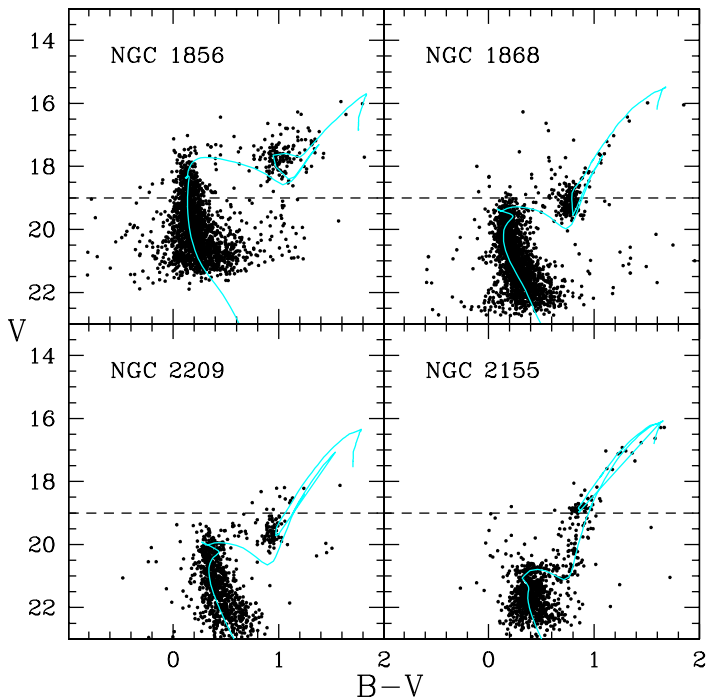} 
 \caption{Examples of observed HST/WFPC2 (V, B--V) CMDs for 4 stellar 
clusters in our sample. These clusters are in a sequence of age, 
from $\sim 0.3$ Gyr (NGC\,1856) to $\sim 3.5$ Gyr (NGC\,2155).
The best solutions found by us using the Padova isochrones 
(\cite[KSB07]{KSB07}) are also shown in the figure. 
}
   \label{fig1}
\end{center}
\end{figure}

{\underline{\it HST/WFPC2 data}}.
We analysed HST/WFPC2 (V, B--V) CMDs for a sample of 15 IACs in the LMC
covering ages from $\sim$0.3 to $\sim$4 Gyr.  
These data come from \cite[Brocato \etal\ (2001)]{Brocato_etal01} and are 
photometrically quite homogeneous, reaching typically V$\sim 22$. 
As can be seen in Fig.\ref{fig1}, these CMDs cover at least 2 
magnitudes along the main sequence (MS) and clearly display 
the core helium burning stars in the red clump (RC).

{\underline{\it CMD modelling}}.
We are modelling these CMDs as SSPs, where the basic steps to generate
a synthetic CMD are the following:
1) we choose the stellar evolutionary model; 
2) we choose an age and metallicity. 
These two steps are equivalent to picking 
up an isochrone from the adopted stellar evolutionary library;
3) we apply our choice of distance modulus and reddening to the
isochrone;
4) we distribute the synthetic 
stars following a Salpeter IMF and a fraction of binaries of 30\%;
5) as a last step we introduce the photometric errors and completeness
as determined from the data. 

{\underline{\it The stellar evolutionary libraries}}.
In Table \ref{tab1} we can see the basic differences in 
the three stellar evolutionary libraries that were used. 
All of them offer models with treatment for convective core overshooting
with similar values for $\Lambda_{OV}$. However, only PEL and BaSTI 
also offer models without overshooting -- the classical or canonical models. 
Another important difference between the libraries 
is their discreteness or
the number of steps in age or metallicity.
In this aspect Padova offers a significant larger number 
of possibilities than the other libraries, specially PEL, where 
there are only three values of metallicity available. 

\begin{table}
  \begin{center}
  \caption{The adopted stellar evolutionary libraries}
  \label{tab1}
 {\scriptsize
  \begin{tabular}{|l|c|c|c|}\hline 
{\bf } & {\bf Padova} & {\bf PEL (or Pisa)} & {\bf BaSTI (or Teramo)} \\ \hline
{\bf Convective core:} & & &  \\ 
{\bf Classical ($\Lambda_{\rm OV}/H_{\rm p}=0$}) ? & no & yes & yes  \\ 
{\bf Overshooting $\Lambda_{\rm OV}/H_{\rm p}\ne 0$} & $\sim 0.25$ & 0.25 & 0.25  \\ \hline
$Z$ (0.001--$Z_{\odot}$) & 0.001, 0.002, 0.004, 0.006 & 0.001, 0.004, 0.008 & 0.001, 0.002, 0.004, 0.008\\
 & 0.008, 0.012, 0.015, 0.019& & 0.010, 0.0198\\ 
 & (8 values) & (3 values) & (6 values) \\ \hline
Age (0.10--4 Gyr) & 33 values & 24 values & 22 values\\ \hline

  \end{tabular}
  }
 \end{center}
\vspace{1mm}
 \scriptsize{
}
\end{table}

{\underline{\it Statistical comparisons}}.
To determine the best models we applied a statistical tool 
to compare simultaneously the observed and synthetic MS and RC.
For the MS we compared the differences in colour along the 
observed and theoretical fiducial lines of each cluster 
using a $\chi^{2}$ statistics to assess their similarity. For
the evolved stars we computed a CMD distance, $\delta_{\rm RC}$, 
between the median positions of real and artificial RC stars. 
So, the best models are the ones that have the minimum values 
in both statistics.

\section{Self-consistent physical parameters}

Table \ref{tab2} summarises the results. The first line
lists the typical formal random uncertainties in each parameter. 
For more details on how these uncertainties are estimated we refer
to \cite[KSB07]{KSB07}.
In the remaining lines, the first (second) number in each entry is the 
systematic (rms) difference in the specific comparison being made.
We compared our previous results based on Padova with the literature,
as well as the distinct evolutionary models among themselves. 
We detail these comparisons below. 

\begin{figure}[b]
\begin{center}
 \includegraphics[width=10.0cm]{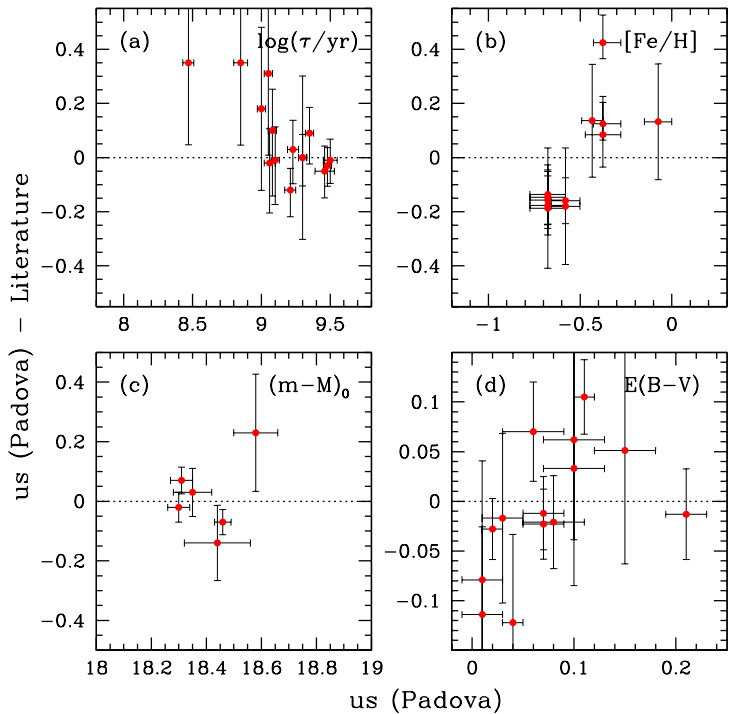} 
 \caption{Differences in the physical parameters
determined by us (using the Padova isochrones) in relation to the ones found 
in the literature (see text for details). }
   \label{fig2}
\end{center}
\end{figure}

{\underline{\it Padova vs. liturature}}. 
Figure \ref{fig2} shows the comparisons between our results 
using the Padova isochrones (\cite[KSB07]{KSB07}) 
with the ones found in the literature. 
We identify an underestimate in age in the previous works, which
were based on ground-based observations 
(e.g. \cite[Elson \& Fall 1988]{EF88}). 
The amplitude of the effect is $\sim 0.30$ in log(age) 
for clusters younger than 1 Gyr. 
On the other hand, for the clusters older than this age limit,
all of them analysed with high quality data (HST or VLT)
by other authors (e.g. \cite[Rich \etal\ 2001]{Rich_etal01}), 
this difference seems disappear.
  
A very interesting comparison can be done based on spectroscopy
of red giants, in particular using the analysis of Calcium triplet 
(\cite[Olszewski \etal\ 1991]{Oetal91}; 
\cite[Grocholski \etal\ 2006]{Grocholski_etal06}).
These comparisons reveal that, despite the scattering of 0.20,
our CMD analysis predicts metallicities without systematic effect.
So, this scattering can be considered as a realistic random 
uncertainty in our method based on a pure CMD analysis. 

Concerning the distance modulus, another interesting comparison 
can be done with the results based on the RC in the K band as a 
distance indicator (\cite[Grocholski \etal\ 2007]{Grocholski_etal07}). 
For the six stellar clusters in common, despite the significant scattering
found, the systematic differences are very small. 
The same can be said in relation to the reddening, where in this case 
the literature results come from analysis of integrated optical colours 
(\cite[McLaughlin \& van der Marel 2005]{MvdM05}).

{\underline{\it Classical vs. Overshooting}}.
As expected (\cite[Gallart, Zoccali \& Aparicio 2005]{GZA05}), 
inside the same library, models with treatment for the 
convective core overshooting recovered older ages in relation to the 
ones obtained by the classical models.
Despite the systematic differences in log(age) of 0.08 
for the PEL models and 0.13 for the BaSTI models,
there is no clear trend in the other parameters.      

{\underline{\it PEL or BaSTI vs. Padova}}. 
The comparisons involving different libraries with 
convective core overshooting reveal that
the adoption of a specific stellar evolutionary model can 
produce a significant bias in the physical parameters. 
Taking the Padova library as a reference, 
the PEL library systematically overestimates distance modulus
by 0.12 (or $\sim 2.8$ kpc at 50 kpc), whereas the BaSTI models 
systematically overestimate ages by 0.12 in log(age) 
($\sim 0.30$ Gyr for an age of 1 Gyr).
Apart from these, no other trend was observed
in the other parameters. 

\begin{table}
  \begin{center}
  \caption{Self-consistent physical parameters. First line: 
typical formal uncertainties. Other lines: systematic and {\it rms} 
differences between the results from different stellar evolutionary 
models.}
  \label{tab2}
 {\scriptsize
  \begin{tabular}{|l|c|c|c|c|}\hline 
 & log(age/yr) & [Fe/H] & (m-M)$_{0}$ & E(B-V) \\ \hline
typical formal uncertainties  & 0.05 & 0.10 & 0.08 & 0.02 \\ \hline
Padova -- literature & +0.30 / 0.10$^1$ & +0.02 / 0.20  & -0.02 / 0.13 & -0.01 / 0.07 \\
 & +0.02 / 0.11$^2$ &  & & \\ \hline
PEL & & & & \\ 
overshooting -- classical & +0.08 / +0.09 & 0.06 / 0.11  & +0.01 / 0.10 & -0.02 / 0.03 \\ \hline
BaSTI & & & & \\
overshooting -- classical & +0.13 / +0.09 & 0.02 / 0.20 & +0.04 / 0.14 & 0.00 / 0.04\\ \hline
PEL (over) -- Padova & +0.00 / +0.06 & +0.04 / 0.17 & +0.12 / 0.11 & 0.00 / 0.04 \\ \hline
BaSTI (over) -- Padova & +0.12 / 0.05 & +0.06 / 0.12 & +0.03 / 0.05 & -0.01 / 0.03 \\ \hline
   \end{tabular}
 }
 \end{center}
\vspace{1mm}
 \scriptsize{
 {\it Notes:} $^1$ age $<$ 1 Gyr ; $^2$ age $>$ 1 Gyr \\
}
\end{table}

\section{The quality of the fit}

\begin{figure}[b]
\begin{center}
 \includegraphics[width=8.5cm]{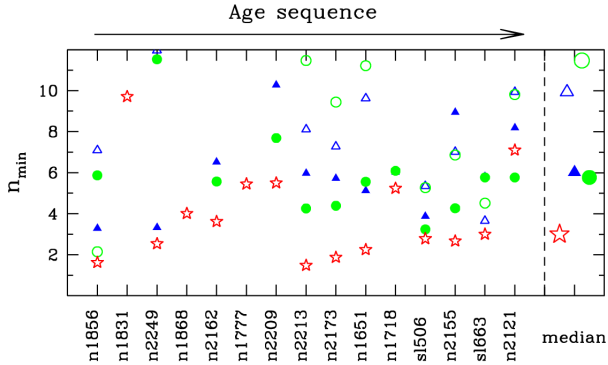} 
 \caption{The quality of the fit for all clusters 
as determined by the $n_{\rm min}$. 
The clusters are ordered according to age, with the youngest
on the left.
The symbols code the different choices of stellar evolutionary models: 
Padova (stars), PEL with (solid triangles) and without (open triangles)
overshooting, BaSTI with (solid circles) and without (open circles)
overshooting. 
The median values for all clusters are also shown in the figure.
}
   \label{fig3}
\end{center}
\end{figure}

To determine the quality of the fit we took simultaneously the MS 
and RC fit into account. We define a parameter called {\it n} 
for each model as 

$$ n = \sqrt{(\chi^{2}/\sigma_{\chi})^{2} + 
(\delta_{RC}/\sigma_{\delta})^{2}}~, $$

where $\sigma$ means the standard deviation in each statistics
as determined by control experiments. 
So, the best model is the one that minimizes this parameter 
($n_{\rm min}$).

As can be seen in Fig. \ref{fig3}, there is no clear age dependence
in the quality of the fit. 
On the other hand, on average, models with overshooting fit better 
the data than the classical ones. 
Furthermore, the quality of the fit is similar for the PEL and BaSTI 
libraries, but significant higher for the Padova models.

\begin{figure}[b]
\begin{center}
 \includegraphics[width=7.5cm]{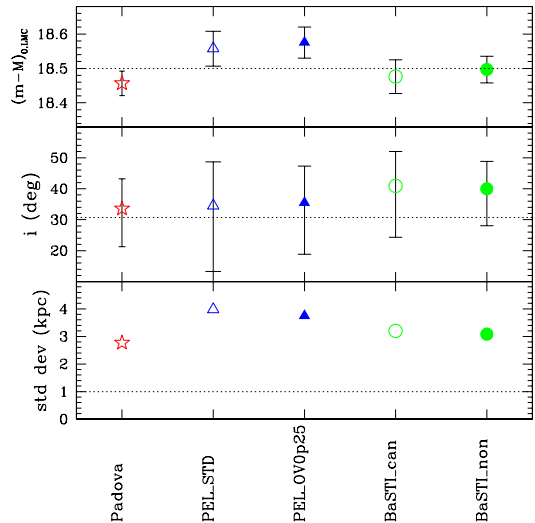} 
 \caption{LMC distance (upper panel), disk inclination (middle panel)
and standard deviation (lower panel) in the fitted 3D distribution
for our sample of 15 stellar clusters. 
 }
   \label{fig4}
\end{center}
\end{figure}

\section{3D distribution and LMC distance}
Since we determined the individual distances to each cluster 
we were able to probe the 3D distribution of these objects and
also to provide new determinations of the distance to the LMC centre.
These results are summarized in Fig. \ref{fig4}, which reveals that:
1) independent of the adopted stellar evolutionary library, 
the IACs have a spatial distribution roughly 
aligned (disk inclination $\sim 35$ deg) with the LMC disk 
as defined by field stars 
(in this case Cepheid stars, \cite[Nikolaev \etal\ 2004]{Netal04});
2) the IACs seem to belong to a thick disk, and the 
spatial scatter around the best-fit geometry is
slightly smaller for models with overshooting;
3) the determined distance modulli to the LMC center 
are marginally consistent with the canonical value of 18.50
(\cite[Clementini \etal\ 2003]{Cetal03});
the PEL models recovered systematically larger values than
the ones obtained using the other libraries.


\begin{thebibliography}{}

\bibitem[Bica \etal (2008)]{Bica_etal08} 
Bica, E., Bonatto, C., Dutra, C. M., Santos, J. F. C. 2008, \textit{MNRAS}, 
389, 678

\bibitem[Brocato \etal\ (2001)]{Brocato_etal01} 
Brocato, E., Di Carlo, E. \& Menna, G. 2001, \textit{A\&A}, 374, 523

\bibitem[Castellani \etal\ (2003)]{Castellani_etal03} 
Castellani, V., Degl'Innocenti, S., Marconi, M., Prada Moroni, P. G., 
\& Sestito, P. 2003, \textit{A\&A}, 404, 465

\bibitem[Clementini \etal\ (2003)]{Cetal03}
Clementini, G., Gratton, R., Bragaglia, A., \etal\ 2003, \textit{AJ}, 125, 1309

\bibitem[Elson \& Fall (1988)]{EF88}
Elson, R. A., \& Fall, S. M. 1988, \textit{AJ}, 96, 1383

\bibitem[Gallart, Zoccali \& Aparicio (2005)]{GZA05}
Gallart, C., Zoccali, M., \& Aparicio, A. 2005, \textit{ARA\&A}, 43, 387

\bibitem[Girardi \etal\ (2002)]{Girardi_etal02} 
Girardi, L., Bertelli, G., Bressan, A., \etal\ 2002, \textit{A\&A}, 391, 195

\bibitem[Grocholski \etal\ (2006)]{Grocholski_etal06}
Grocholski, A.J., Cole, A.A., Sarajedini, A., Geisler, D. \& Smith, V.V.
2006, \textit{AJ}, 132, 1630 

\bibitem[Grocholski \etal\ (2007)]{Grocholski_etal07}
Grocholski, A.J., Sarajedini, A., Olsen, K., Tiede, G., Mancone, C. 
2007, \textit{AJ}, 134, 680

\bibitem[Kerber \etal\ (2007)]{KSB07} 
Kerber, L., Santiago, B. \& Brocato, E., 2007, \textit{A\&A}, 462, 139 (KSB07)

\bibitem[McLaughlin \& van der Marel (2005)]{MvdM05}
McLaughlin, D. E., \& van der Marel, R. P., 2005, \textit{AJ}, 161, 304

\bibitem[Nikolaev \etal\ (2004)]{Netal04}
Nikolaev, S., Drake, A.J., Keller, S.C., \etal\ 2004, \textit{ApJ}, 601, 260

\bibitem[Olszewski \etal\ (1991)]{Oetal91} 
Olszewski, E.W., Schommer, R.A., Suntzeff, .B. \& Harris, H. 1991, 
\textit{AJ}, 101, 515

\bibitem[Pietrinferni \etal\ (2004)]{Pietrinferni_etal_04} 
Pietrinferni, A., Cassisi, S., Salaris, M., Castelli, F. 2004, 
\textit{ApJ}, 612, 168

\bibitem[Rich \etal\ (2001)]{Rich_etal01}
Rich, R. M., Shara, M. M., \& Zurek, D. 2001, \textit{AJ}, 122, 842

\end{thebibliography}
\end{document}